# Modules in the metabolic network of *E.coli* with regulatory interactions

Jan Geryk[a], František Slanina[b]

[a] *Department of Philosophy and History of Sciences, Faculty of Science, Charles university, Viničná 7,128 44 Prague, Czech Republic*

[b] *Institute of Physics, Academy of Sciences of the Czech Republic, Na Slovance 2, CZ-18221 Praha 8, Czech Republic*

e-mail adresses: *geryk.cz@gmail.com (Jan Geryk), slanina@fzu.cz (František Slanina)*

**Abstract**

We examine the modular structure of the metabolic network when combined with the regulatory network representing direct regulation of enzymes by small metabolites in *E.coli*. In order to identify the modular structure we introduce clustering algorithm based on a novel vertex similarity measure for bipartite graphs. We also apply a standard module identification method based on simulated annealing. Both methods identify the same modular core each of them with different resolution. We observe slight but still statistically significant increase of modularity after regulatory interactions addition. Enrichment of the metabolic network with the regulatory information leads to identification of new functional modules, which cannot be detected in the metabolic network only. Regulatory loops in the modules are the source of their self-control, i.e. autonomy, and allow to make hypothesis about module function. This study demonstrates that incorporation of regulatory information is important component in defining functional units of the metabolic network.

# 1. Introduction

One of the most studied properties of the real networks is their modularity. The idea of modularity is widely accepted in diverse fields (neurophysiology, computer science, evolutionary biology, etc.). In this context, a module represents a relatively autonomous system with an elementary function. It remains a challenging problem to find cellular modules solely on the basis of the network topology representing molecular interactions within the cell. We can expect auto-regulation and robustness in the functional modules. In the graph model, these properties are represented by high density of edges inside modules. Relative autonomy of modules implies low density of edges between modules in the graph representation. In the graph theory, the module identification is transformed into the question of how to find a partition of a graph with maximum density of edges inside subgraphs and minimum density of edges between subgraphs. There are a number of methods that solve the question and provide efficient algorithms for detection of modules in the networks , but open questions still remain (Barber, 2007; Ding et al., 2006, Fortunato, 2010; Guimera, Sales-Pardo, and Amaral, 2008; Lancichinetti and Fortunato, 2009; Newman, 2004; Newman and Girvan, 2004; Palla et al., 2005; Rosvall and Bergstrom, 2007, Zhang et al.,2009).

One of the studied problems in the field of metabolic network research is the distribution of classical metabolic pathways among the modules. These metabolic pathways are defined on the basis of biochemical knowledge and are accessible in the KEGG database (Kyoto encyclopedia of genes and genomes). It was demonstrated that one metabolic pathway is typically distributed among more than just one module at the same time. Within the module, there is typically more than one metabolic pathway (Guimera and Amaral, 2005; Zhao et al., 2006). These results show that it is impossible to assign  the known metabolic pathways unambiguously to modules identified on the basis of the network topology. One may hypothesize that topological modules have specific functions that cannot be satisfactorily captured by classical metabolic pathway categorization. However, this hypothesis has never been tested. Several analyses performed during the last few years are in accord with the hypothesis of evolutionary autonomy of modules. It was confirmed that the modularity measure depends on the variability of bacteria's live environment. Bacterial strains living in variable (unpredictable) environments have more modular metabolic networks in comparison with strains that live under constant environmental conditions (Kreimer et al., 2008; Parter, Kashtan, and Alon, 2007). Moreover, Alon and Kashtan (2005) predicted these findings by modeling the evolution of boolean networks. It was also shown that enzymes within a topological module have tendency to co-occur in the set of metabolic networks of 54 taxa implying evolutionary conservation of modules (Zhao et al., 2007). But from another point of view, modules or network partitions obtained solely on the basis of reaction co-occurrence within phylogenetic system (Wagner, 2009) were not systematically compared with topological modules. In yet another study Guimerá and Amaral (2005) show that non-hub nodes (metabolites) which provide interface between modules are evolutionary more conserved than the rest of network nodes. The current knowledge can offer several indications supporting relevancy of metabolic network's modular structure but functional interpretation of detected modules is still insufficient or missing.

This paper is focused on the modular structure of bipartite representation of *E. coli* metabolic network. A standard metabolic network representation is unipartite, .i.e. the network has a single type of nodes only - typically the metabolites (Ma and Zeng, 2003a; Ma and Zeng, 2003b). In the bipartite representation used in this work, two types of nodes are present - metabolites and enzymes. Edges can be placed only between metabolites and enzymes. Bipartite representation allows to integrate the regulatory interactions together with the metabolic network in a straightforward way. The effect of the addition of the regulatory interactions on the modular structure is especially analyzed. Assumption of the functional autonomy of modules implies their auto-regulation. We hypothesize that regulatory interactions are concentrated in the functional modules. We compare the modular structures identified within the metabolic network with and without regulatory interactions from the quantitative and qualitative point of view.

## 2. Methods
### 2.1 Data extraction

The dataset from a previously published paper, where the metabolic network of *E. coli* was reconstructed using EcoCyc 9.0 database is used (Seshasayee et al., 2009). The complete list of removed currency metabolites is available in the paper mentioned above. We constructed the bipartite graph where one set of nodes are metabolites and the second one are enzyme genes using this dataset. An edge is placed between the metabolite and the enzymatic gene if the metabolite is a substrate or product of the enzyme coded by this gene. As a second step we reduced the complexity of the network by replacing every set of enzymatic genes with same neighbors by one node. Enzymatic genes with same neighbors correspond in most cases to an enzymatic complex which catalyses one reaction or to isoenzymes. In the bipartite graph representation they form complete subgraphs which are expected to be in the centers of modules. We avoid the impact of these complete subgraphs on the detected modular structure by representing genes with the same neighbors by a single vertex. For the subsequent analysis, the largest connected component from the reconstructed bipartite graph is used.

Regulatory interactions were extracted from the EcoCyc 9.0 database, particularly from the file „regulation". The metabolites in this file are assigned to reactions which they regulate. With another file ("reaction") from the same database it is possible to assign enzymes (or EC numbers respectively) to reactions. The enzymes are represented as Blattner-ID of their corresponding genes in the dataset of Seshasayee et al. (2009). In order to assign EC nubres to Blattner ID's the "eco_enzyme.list" file from the KEGG database was used.

Extracted regulatory interactions are then placed into the bipartite graph in the form of additional edges. If non-metabolite node corresponds to the set of enzymatic genes with same neighbors a regulatory edge is placed between the metabolite and the non-metabolic node if the metabolite regulates at least one enzyme or enzyme subunit coded by some of these genes. By this procedure, the metabolic network combined with the regulatory network is constructed. In the following we talk about enzymes or enzyme nodes to mean non-metabolite nodes in our bipartite representation.

**2.2 The module identification algorithms**

The procedure is centered on the quantity measuring local density of edges (vertex similarity measure). The portions of the graph where this quantity is larger are more likely to belong to the inside of modules. Let us denote $E(u,v)$ the number of edges within the induced subgraph that is determined by the neighbors of nodes $u$ and $v$, $u \in U$ ($U$ is the set of metabolites), $v \in V$ ($V$ is the set of reactions), $k_v$ is the number of node $v$ neighbors and $k_u$ is the number of node $u$ neighbors. In the following we are concentrated only on the local density of edges in the neighborhood of two vertices connected by the edge, so the simplest definition of the local density of edges would be:

$$\tau(u,v) = \frac{E(u,v)}{k_u k_v}. \qquad (1)$$

However, this definition has a serious drawback. The local density defined by eq. (1) attains relatively high values if the induced subgraph of the connected nodes $\{u, v\}$ is a small tree. The maximum value of $\tau$ is attained in the case of star with arbitrary size. The tree structures do not correspond to intuitive idea of modules. Therefore, throughout this work we use the following definition of the local density of edges in the neighborhood of two connected vertices.

$$\tau(u,v) = \frac{E(u,v) - k_v - k_u + 1}{(k_u - 1)(k_v - 1)}. \qquad (2)$$

For $k_v = 1$ and/or $k_u = 1$, $\tau(u,v) = 0$ is defined. From the above definition becomes clear that for any tree subgraph will be $\tau(u,v) = 0$. In order to measure density of edges in the identified modules we use formally same equation as equation 2. Let us denote number of edges inside module $s$ $E_s$, number of metabolites in $s$ $nu_s$ and number of enzymes in $s$ $nv_s$. The normalized density of module $s$ is defined as follows:

$$D_s = \frac{E_s - nu_s - nv_s + 1}{(nu_s - 1)(nv_s - 1)}. \qquad (3)$$

For $nu_s = 1$ and/or $nv_s = 1$, $D_s = 0$ is defined as in previous case. The mean of $D_s$ over all modules is denoted by $D$.

As in several other procedures to find modules, in the course of our algorithm we shall need a measure to quantify how good is that partitioning of vertices among modules. To this end we use the standard modularity measure (Newman and Girvan, 2004), with a slight modification, in order to take into account the bipartite character of the network. The modification is explained in detail in Appendix A. Therefore, we define.

$$Q^B = \frac{1}{L} \sum_{s=1}^{m} \left( l_s - \frac{d_s h_s}{L} \right) \qquad (4)$$

where $L$ is the number of edges in the bipartite graph, $l_s$ is the number of edges inside the module $s$, $m$ is the number of modules in the bipartite graph, $d_s$ is the sum of metabolite degrees in the module $s$ and $h_s$ is the sum of reaction degrees in the module $s$. Modularity measure is the difference between the number of edges inside modules and the expected value of this quantity inside a random graph ensemble with the same degree sequence as in the original graph.

Our algorithm for finding modules in bipartite graph is based on the idea that edges with higher $\tau$ are more likely to be placed within the modules. In some sense it is an inverse procedure to the algorithm used in (Newman and Girvan, 2004) and its variants. The algorithm starts with the bare set of vertices and no edges. We add edges one by one, starting with the edge with largest $\tau$ and continuing in the order of decreasing $\tau$. If more than one edge has the same value of $\tau$, all of them are placed at once. At each step, we obtain a graph composed of one or more components representing potential modules. For the partitioning we obtained we calculate the modified modularity measure (3). In the course of the algorithm, the value of $Q^B$ evolves. For the subsequent analysis, we use the modules which emerged at such step, in which $Q^B$ attained maximum value.

To compare our method with the mainstream module detection method we also applied the simulated annealing module identification method to the studied metabolic network. The simulated annealing for module identification is a stochastic optimization method where the optimized quantity is modularity measure $Q$ (Guimera and Amaral, 2005). The procedure starts with arbitrary partition ($A$) of the network. In the next step neighboring partition ($B$) of the starting partition is generated, typically by moving one node from one module to another module and modularity measure $Q(B)$ for the newly generated partition is computed. If $Q(A) \leq Q(B)$, the partition $B$ is accepted as a new starting partition. If $Q(A) > Q(B)$, the partition $B$ is accepted with probability $p = \exp\left(-\frac{Q(A)-Q(B)}{T}\right)$. $T$ is a parameter that controls the probability of accepting partitions with decreasing modularity. During the procedure, $T$ is continuously decreasing. This allows broader search of partition space at the beginning, continues to be more stringent and results to $p\sim 0$ in the last steps of the procedure. Modules from the last partition with highest $Q$ are considered as relevant modules of the network.

**2.3 Significance of maximum modularity value of the network**

Randomization method described in (Maslov, Sneppen, and Zaliznyak, 2004) is used to assess the significance of the maximum modularity value. The principle of the method is to apply local randomization repeatedly in the graph. In each local randomization step, two edges $\{a,b\}$ and $\{c,d\}$ are randomly selected, removed from the graph and new edges: $\{a,d\}$ and $\{c,b\}$ are created provided that edges $\{a,d\}$ and/or $\{c,b\}$ are not already present. If edges $\{a,d\}$ and/or $\{c,b\}$ are already present, the random selection of the two edges is repeated until it is possible to swap them.

During randomization of metabolic network the graph connectivity is controlled and only randomizations that conserved the connectivity of the graph are accepted. To obtain one randomized version of the metabolic network 30000 local randomization steps were applied as described above. Sixty randomized metabolic networks were generated and maximum modularity value $\max(Q_{rand}^B)$ was computed for each of them by applying the clustering algorithm. The null hypothesis that the maximum modularity value $\max(Q^B)$ obtained with the original metabolic network is smaller than the random sample from the normal distribution with the expected value and standard deviation computed from the ensemble of 60 randomized networks is tested.

In the case of regulatory network, connectivity constraint during randomization is relaxed because this network is disconnected in itself. As in the previous case, 30000 local randomizations were applied to obtain one randomized regulatory network and 60 randomized regulatory networks was generated in total. Every randomized regulatory network was assembled with the original metabolic network and $\max(Q_{rand}^B)$, was computed by applying the clustering algorithm. With this ensemble it is possible to test the statistical significance of the modularity increase after assembling a metabolic network with a regulatory network. As in the previous case z-test is used to test whether the maximum modularity value obtained with the original metabolic network combined with the original regulatory network is smaller than the random sample from normal distribution with mean and standard deviation computed from the randomized ensemble of regulatory networks combined with nonrandomized metabolic network.

**2.4 Significance of the KEGG category content in the identified modules**

A commonly used model to test statistical significance of functional category content in the module is the hypergeometric distribution. This model does not reflect the way the modules or a network partition were obtained assuming the nodes in the module are sampled quite randomly from the set of network nodes. A typical module detection algorithm implicitly prefers connected subgraphs as modules. In the clustering algorithm that is used in this work, the modules are defined as connected subgraphs in the network partitions obtained by successive reconstruction of the network. The effect of connectedness should be filtered out in order to test the significance of the metabolic category content in the modules.

For each module size obtained by the clustering algorithm, 100000 connected subgraphs (of that size) were randomly sampled from the metabolic network with or without regulatory interactions and the KEGG category distribution in the randomly sampled subgraphs was recorded. For each module identified by the clustering algorithm (section 2.3) and the KEGG category dominant in the module the empirical $p$-value was computed by counting the fraction of randomly sampled connected subgraphs of the corresponding size with larger or equal content of the KEGG category (that is dominant in the identified module). The KEGG categories correspond to 11 general metabolic classes or maps defined on the KEGG webpage.

## 3. Results
### 3.1. Quantitative comparison

We applied our clustering algorithm and simulated annealing method both on the metabolic network without regulatory interactions and on the metabolic network combined with regulatory network. In the second case there are two alternatives how to control the algorithm flow. In the first alternative, the regulatory and reactionary edges are not distinguished and $\tau$ is computed for every edge in the graph. In the second alternative, $\tau$ is computed only for edges that represent reactionary (and not regulatory) relationship between the metabolite and reaction node ensuring the reactionary connectedness of identified modules. We investigate both possibilities.

All quantities we used for comparing mentioned methods and modular structures identified in the network with and without regulations, are summarized in the table 1. The main difference between both module identification methods is the value of modularity maximum $\max(Q)$. The reason is that not all network edges are partitioned in to the modules after applying our clustering method. This is due to constraint of $\tau$ that determine the way modules are constructed. There is no constraint in the local density of edges in case of the simulated annealing method. As a result, all edges are partitioned into the modules, increasing number of positive summands in the equation 3.

Consider a network with dense core subgraph and sparse rest of the network, the periphery. Even if periphery of the network will be absolutely non-modular (for example created by one linear chain of nodes) it may have a relatively high value of modularity for many possible partitions.

We show this more precisely in supplementary materials. This idealized situation is similar to our results on metabolic network. The clustering method identifies core modules leaving the rest of the network non-partitioned. The simulated annealing identifies similar core modules but also many other modules with very low edge density. These low edge modules are source of higher modularity approached by simulated annealing.

In order to prove this proposition quantitatively we use two partitions of the metabolic network combined with regulatory network. The first partition is generated by the simulated annealing and the second partition by clustering method without constraint of reactionary connectedness. Results obtained by using partitions produced by clustering algorithm with imposed reactionary connectedness and results obtained by using metabolic network without regulations are similar. First we substract all nodes not contained in the modules identified by the clustering method from the partition produced by the simulated annealing. Thus, we obtain reduced partition $P_r$ which divides into the modules the same subset of network nodes like clustering method. The normalized mutual information (Guimerà et al., 2006) between the partition $P_r$ and partition produced by clustering algorithm is $I_{norm} = 0.814$, a value confirming relatively high similarity. The modularity of $P_r$ is 0.378, which is very close to one obtained by clustering algorithm (0.381). The modularity of remaining modules not contained in $P_r$ is 0.142. In total, we got modularity 0.520 approaching the value 0.604 produced by the simulated annealing. (The difference 0.84 is due to fact that some of the modules are broken by the division of the partition generated by simulated annealing, according to partition generated by clustering algorithm). We also compute a mean of normalized density ($D_{core}$) of modules defined by the partition $P_r$ and the same quantity denoted ($D_{periphery}$) for the rest of modules not contained in $P_r$. We obtain $D_{core} = 0.388$ and $D_{periphery} = 0.0029$. For the partition generated by the clustering algorithm we obtain $D = 0.271$. These results confirm high-density modular core identified by both methods and the sparse non-modular periphery partitioned only by the simulated annealing method. The same fact is also reflected by the mean density of modules ($D$) produced by simulated annealing which is significantly smaller than the same quantity produced by clustering algorithm in all considered variants (tab.2).

The main difference between both methods in terms of partitioning of the core is the resolution level modules are detected on. The simulated annealing tends to generate smaller modules than clustering method. Some of the core modules detected by the clustering algorithm are divided into smaller modules by the simulated annealing.

The observed increase of modularity after addition of regulatory edges is significant on the basis of z-test ($p < 0.01$). Combination of the randomized regulatory network with the non-randomized metabolic network leads to the modularity decrease in average (tab.1). The effect of modularity increase is not observed after applying simulated annealing method. If we reduce the partition produced by the simulated annealing (applied on the metabolic network without regulations) to the nodes contained in the partition obtained by the clustering method and compute modularity we obtain value 0.293. The same procedure using metabolic network combined with regulatory interactions leads to the value 0.378, implying that the effect of modularity increase is localized in the modular core.

The analysis of KEGG category content shows a weaker consistency of identified modules with traditional partitioning of metabolism into the functional units. Both quantities used are similar for both applied methods as well as for metabolic network with and without regulatory interactions (tab.1).

Tab.1.

|  | $D$ | $\max(Q^B)$ | mean/std of $\max(Q^B_{rand})$ | mean of max. fraction of nodes in one KEGG category | % of modules with significant KEGG category content |
|---|---|---|---|---|---|
| **Metabolic network without regulatory interactions** | | | | | |
| Clustering | 0.422 | 0.310 | 0.084/0.007 | 0.681 | 38 |
| Simulated anealing | 0.210 | 0.658 | - | 0.635 | 36 |
| **Metabolic network with regulatory interactions** | | | | | |
| Clustering | 0.271 | 0.381 | 0.231/0.010 | 0,673 | 37 |
| Clustering with reactionary connectedness | 0.309 | 0.341 | 0.233/0.006 | 0.685 | 36 |
| Simulated anealing | 0.181 | 0.604 | - | 0.657 | 38 |

**3.2. The biochemical structure of modular core**

During comparison of partitions produced by the module identification algorithms used in this work, we recognized four types of modules.

Modules of first type exhibit a high density of reactionary edges and are identified by both considered methods and in both cases with and without regulatory interactions. Typical example of this type is module corresponding to the metabolism of vitamin B6. This module corresponds to module 1.4 (fig.1). and to module 2.3 (fig.2 and 3). Second example is module corresponding to the synthesis of s-adenosyl-L-homocysteine from the L-methionine. This module corresponds to module 1.2 (fig.1). and to module 2.2.2 (fig.2 and 3).

Modules of second type are significantly similar in both methods and in comparison of the networks with and without regulatory interactions. These modules are typically divided into small number of dense submodules by the simulated annealing. The example is the biggest identified module corresponding to metabolism of nucleotides. This module corresponds to module 1.5 (fig.1) and module 2.5 (fig.2).

Modules of first and second type cause the significant similarity between the partitions of the metabolic network without regulatory interactions and partitions of the network with regulatory interactions.

Modules of third type are identified by both methods only in case of metabolic network combined with regulatory interactions. A typical example is module 2.1 (fig.2 and 3) corresponding to the synthesis of activated forms of glucose from the maltotetraose. Second example is module 2.4 (fig.2 and 3) corresponding to linear synthesis pathway of D-glucuronate from D-galacturonate.

Modules of fourth type are recognizable only in case of metabolic network combined with regulatory interactions. These modules are divided into small number of dense modules by the simulated annealing. Typical examples are module 2.2.1 (fig. 2 and 3) corresponding to the part of glycolysis were fructose-1,6-bisphosphate is cleaved to dihydroxyacetone phosphate and glyceraldehyde-3-phosphate and module 2.2.3 (fig. 2 and 3) corresponding to metabolism of aminoacids.

Modules of third and fourth type correspond to the sparse tree structures in the very metabolic network. After addition of regulatory edges, they became denser and so detectable by the module identification algorithms. In the metabolic network without regulatory interactions, these modules look like arbitrary or random parts of the network and there is no reason why would they be relatively autonomous functional units. We argue that the regulatory loops contained in these modules are sources of autonomy and functional interpretability.

Let us concentrate on the module 2.2.1. The module contains important glycolytic reactions, especially fructose-1,6-bisphosphate cleavage to dihydroxyacetone phosphate and glyceraldehyde-3-phosphate. There is also cleavage reaction of tagatose-1,6-bisphosphate to the same products and reaction converting dihydroxyacetone phosphate to glycerol phosphate. If we consider regulatory interactions, this system become closed due to many regulatory loops and physiologicaly interpretable. Concentration of dihydroxyaceton phosphate and glyceraldehyd 3-phosphate is elevated during glycolysis activation and the regulatory effect they provide within the module is pronounced. Both molecules inhibit their alternative utilization in the pahways other than glycolysis. In the same time fructose 1,6 phosphate inhibits generation of dihydroxyaceton phosphate and glyceraldehyd 3-phosphate from the alternative sources. Phosphoenol pyruvate is also elevated when flux through glycolysis increases. Phospho-enol pyruvate inhibits alternative utilization of dihydroxyaceton phosphate and glyceraldehyd 3-phosphate too and in addition facilitates glycolysis by activation of 6-phosphofructokinase and fructose bisphosphate aldolase. We propose that the function of this module is to fix and facilitate metabolic flux through glycolysis when glycolysis is elevated, for example in the environment with increased concentration of glucose. In principle, the investigated module represents sophisticated positive feedback regulation of glycolysis.

## 4. Summary and discussion

Our results suggest that metabolic network is composed of modular core and non-modular sparse periphery. The similar result was reported by Zhao et al. (2006). In contrast to simulated annealing method our clustering algorithm is capable to selectively identify modular core, leaving the rest of network non-partitioned. We observe statistically significant increase of modularity of the core after addition of regulatory edges. However the modularity difference is very small (0.07). It seems important to perform similar analyses on the different graph representations to make decision about effect of regulatory interactions addition on the network modularity. It is well known fact that modularity of the metabolic network depends strongly on the graph representation. Our bipartite representation leads to sparser graph than the conventionally used unipartite representation. It results to smaller modularity value approached by the simulated annealing.

The metabolism of nucleotides is dominant and significantly abundant category in the biggest identified modules. This is true for both methods and for the metabolic network with and without regulatory interactions. Nucleotides have two crucial functions in the living cell. They are donors of energy in the majority of cellular processes and also precursors of DNA and RNA synthesis. Our analysis shows that metabolism of nucleotides is the most integrated part of the metabolic network of both reactionary and regulatory perspective. This result is in accordance with their crucial importance for the cell.

In the metabolic network without regulations is difficult to interpret identified modules as autonomous functional units. The situation will change after regulatory edges addition. In the section 3.2 we demonstrated that due to regulatory loops within module it is possible to generate hypothesis about module function. The hypothesis about positive feedback system within glycolysis pathway formulated in the section 3.2 is testable by the dynamic modeling but this ambition is out of scope of our article. It was recently shown that it is possible to explain specific experimental behavior of E. coli on the basis of relatively simple metabolic subsystem with regulatory feedbacks. In addition, clearly defined function of the studied subsystem emerged as a consequence of considering regulatory interactions (Kotte et al., 2010).

The relative autonomy of the modules identified in the metabolic network with regulatory interactions is not only in virtue of the sparse connection to the rest of the network implied by the definition of module but also in virtue of auto-regulations included in the modules. This conclusion follows from the obvious idea that system can be autonomous only if it manifests some kind of self-control. The graph representation captures this notion very coarsely but it is important to investigate what we can say about it from the graph perspective.

**Acknowledgements**
J.G. thanks all the members of Department of Philosophy and History of Sciences for inspirative discussions and suggestions. This work was carried out within the project AV0Z10100520 of the Academy of Sciences of Czech Republic and was supported by the MSMT of the Czech Republic, grant no. OC09078 and by the Czech Ministry of education MSM 0021620845.

**Figure captions**

Fig.1. The modular core in the metabolic network without regulatory interactions identified by the clustering algorithm. Enzymes are represented as black nodes. White nodes with black borders represent metabolites.

Fig.2. The modular core in the metabolic network with regulatory interactions identified by the clustering algorithm. Three modules (2.2.1, 2.2.2 and 2.2.3) are detected as one module by clustering algorithm without constraint of reactionary connectedness. In all other cases these modules are detected separately. Enzymes are represented as black nodes. White nodes with black borders represent metabolites. Regulatory interactions are represented as dotted lines

Fig.3. Essential reactions of four modules (2.1, 2.2, 2.3, 2.4) described in the text. Dotted lines without sign represent inhibitory regulations, dotted lines with + sign represent activations.

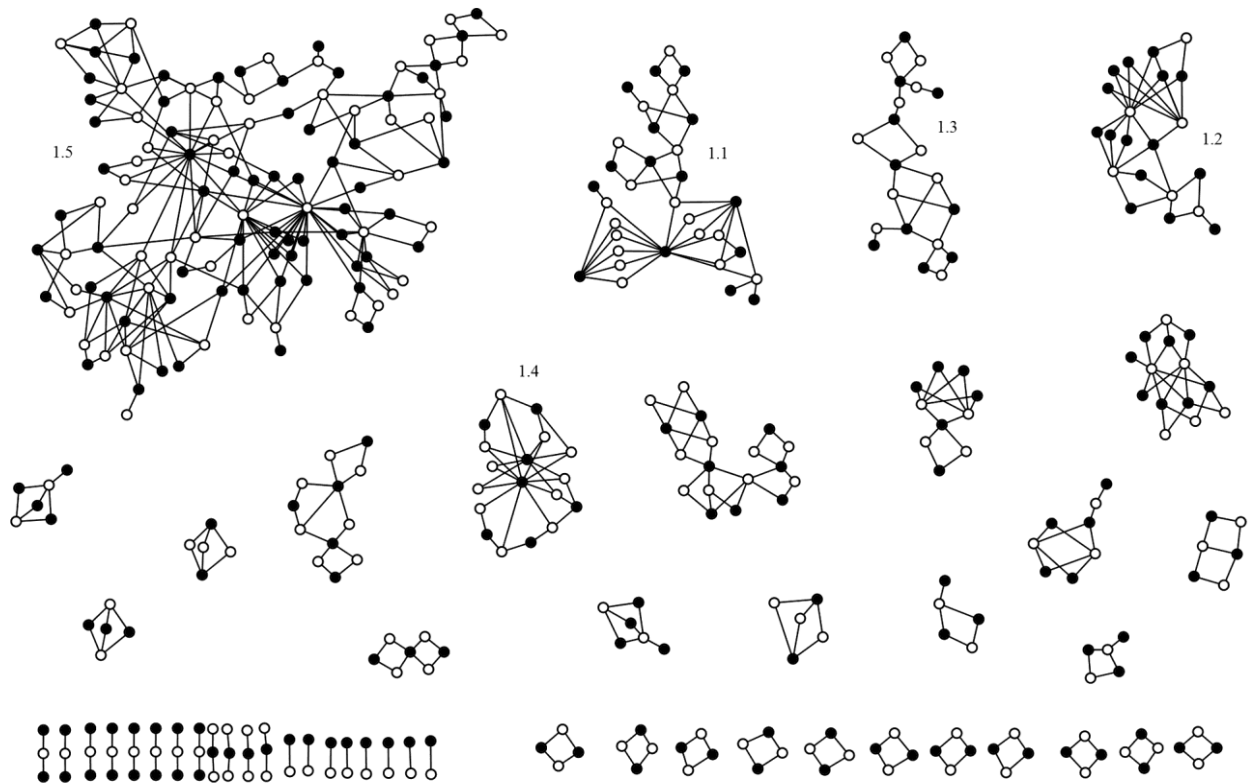

Fig. 1.

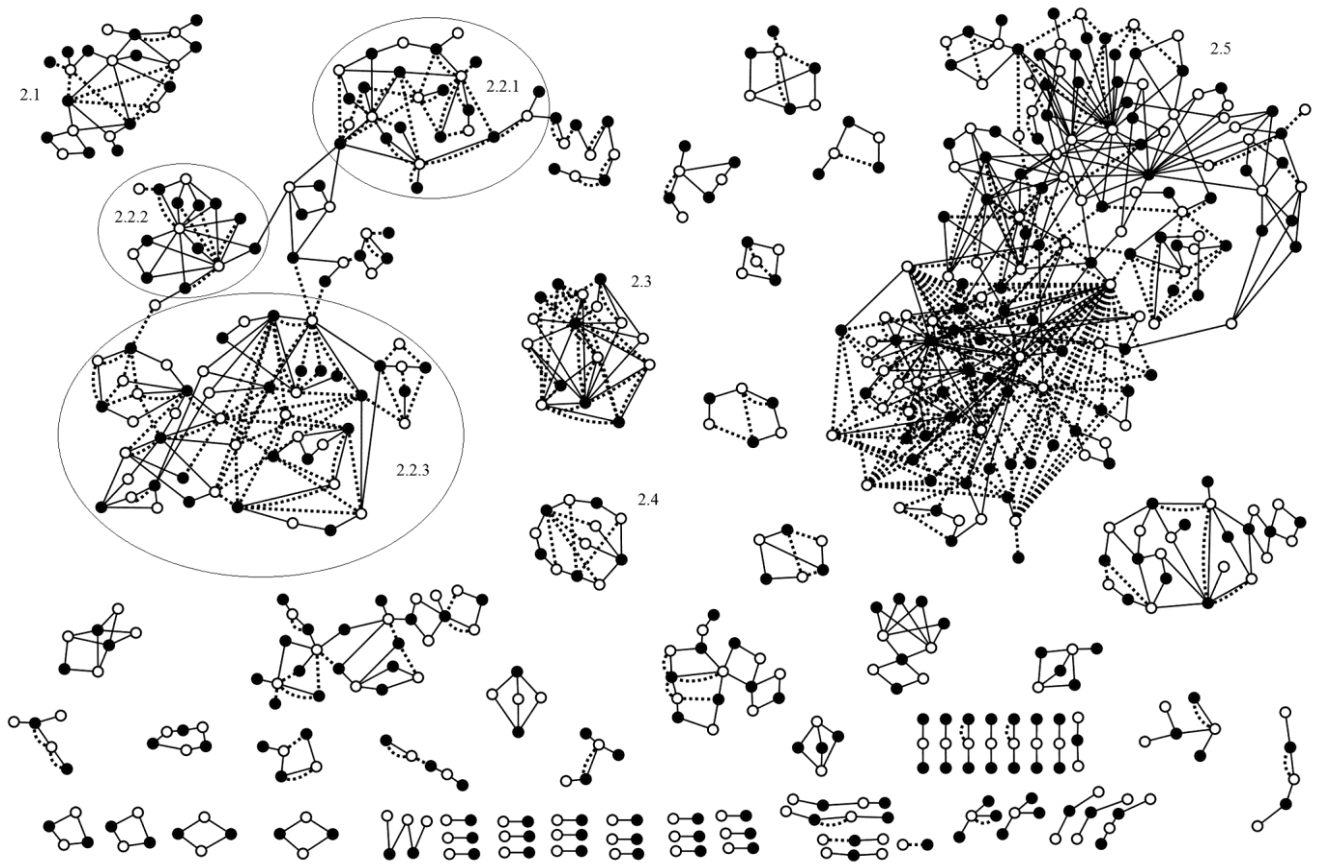

Fig. 2.

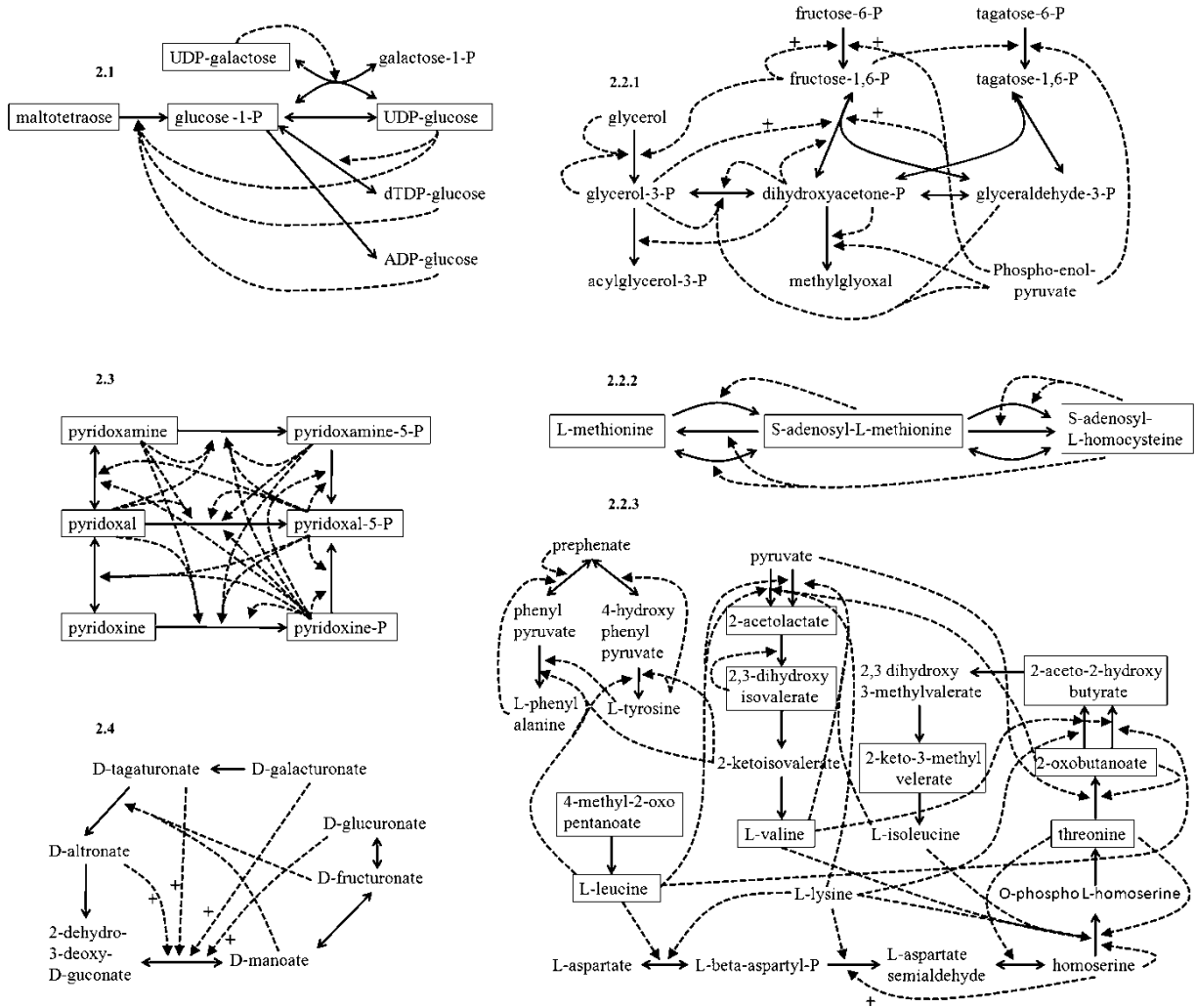

Fig. 3.

**SUPPLEMENTARY FILE**

**Modularity measure for bipartite graph.**
The general formula for the modularity measure is:

$$Q = \frac{1}{L}\sum_{s=1}^{m}[l_s - E(l_s)] \quad \text{s.1}$$

where $L$ is the number of edges in the graph, $m$ is the number of modules, $l_s$ is the number of edges inside the module $s$ and $E(l_s)$ is the expected number of edges between nodes of the module $s$ in the random graph ensemble. We obtain $E(l_s)$ as the sum of probabilities that an edge exist between nodes in the module $s$. In the case of a bipartite graph:

$$E(l_s) = \sum_{u \in U_s, v \in V_s} p(u,v) \quad \text{s.2}$$

where $U_s$ is the set of all metabolites within the module $s$ and $V_s$ is the set of all enzymes within the module $s$. The probability $p(u,v)$ can be interpreted as a number of graphs in the random graph ensemble that contain an edge $\{u,v\}$, divided by the number of all graphs in this ensemble. $p(u,v)$ in the random bipartite graph ensemble with prescribed degree sequence is estimated in the following text. Virtually, we can construct the bipartite graphs from this ensemble by connecting the "stubs" (or half of edges) arising from the metabolites and enzymatic genes. There are $2L$ stubs in the graph, $L$ arising from metabolites and $L$ from reactions. There are $L!$ possibilities how to construct bipartite graph. If vertices $u$ and $v$ are connected by an edge, the number of possibilities how to construct graph is reduced. There are $k_u k_v$ possible realizations of an edge $\{u,v\}$ and after one of these realizations is chosen there is $(L-1)$ number of remaining edges to be placed. The number of possibilities to construct bipartite graph with imposed constraint that between vertices $u$ and $v$ must be an edge is estimated as $k_u k_v (L-1)!$. The probability $p(u,v)$ is estimated as follows:

$$p(u,v) \cong \frac{k_u k_v (L-1)!}{L!} = \frac{k_u k_v}{L} \quad \text{s.3}$$

The same result was obtained in (Barber, 2007). Expected number of edges inside subgraph $s$ is then:

$$E(l_s) \cong \sum_{u \in U_s, v \in V_s} \frac{k_u k_v}{L} = \frac{d_s h_s}{L} \qquad \text{s.4}$$

where $d_s = \sum_{u \in U_s} k_u$ and $h_s = \sum_{v \in V_s} k_v$. With this estimate, it is possible to define modularity measure for bipartite graph:

$$Q^B = \frac{1}{L} \sum_{s=1}^{m} \left( l_s - \frac{d_s h_s}{L} \right) \qquad \text{s.5}$$

**Modularity of sparse periphery**

Consider a partition of the bipartite network on two parts. The first part is arbitrary. The second part is one linear chain of nodes connected by two ends with first part of the network. Let us denote number of nodes in the second part $N_p$ and number of network edges $L$. For simplicity we divide the second part to the $N_p/n$ modules with the same sizes $n$. From the equation s.5 directly follows that the modularity of the second part is:

$$Q_p = \frac{N_p}{L} \left( 1 - \frac{1}{n} - \frac{n}{L} \right) \qquad \text{s.6}$$

If we have fixed values of $N_p$ and $L$ we can still manipulate the value $Q_p$ by choosing the size of modules $n$. We used values $L = 2213$ and $N_p = 516$ corresponding to the application of our clustering algorithm on the metabolic network with regulatory interactions ($N_p$ is the number of nodes not partitioned into the modules by the clustering algorithm). In this case if we choose $= 20$, than $Q_p = 0.22$. This demonstrates that clearly non-modular structures may contribute significantly into the total modularity of network.

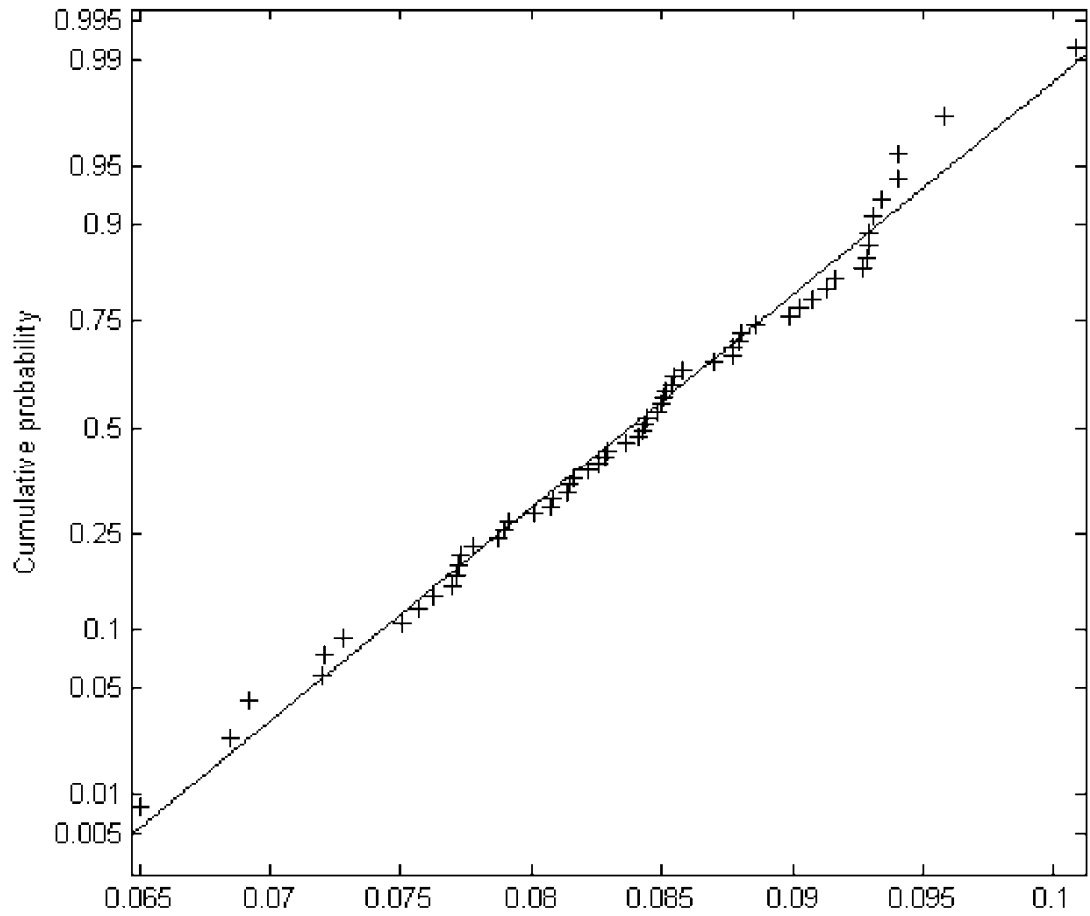

Supplementary Fig.1. The ordered values of $\max(Q_{rand}^B)$ (x-axis) from 60 randomized metabolic networks are plotted against their observed cumulative frequency. The y-axis is scaled for the normal distribution.

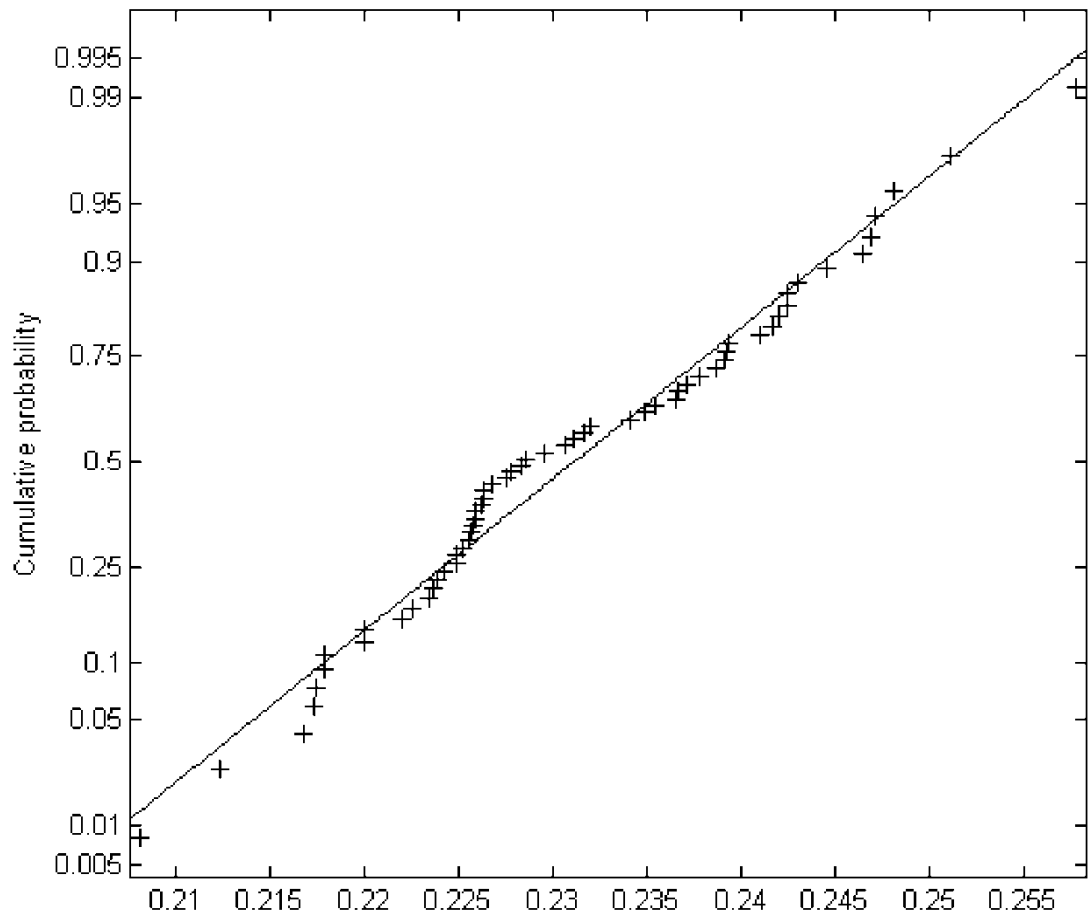

Supplementary Fig.2. The ordered values of max($Q^B_{rand}$) (x-axis) computed from 60 randomized regulatory networks combined with original metabolic network are plotted against their observed cumulative frequency. The y-axis is scaled for the normal distribution.